\documentclass[twocolumn]{aastex63}
\usepackage{lineno}

\shorttitle{Simulating Light Curves of Binary Stars Passing through Micro-Caustics in Strong Lensing Galaxy Clusters}
\shortauthors{Wenwen Zheng et al.}

\graphicspath{{./}{figures/}}

\usepackage{amsmath}	
\usepackage{amssymb}

\begin{document}
\begin{sloppypar}

\title{The Magnified Waltz: Simulating Light Curves of Binary Stars Passing through Micro-Caustics in Strong
 Lensing Galaxy Clusters}

\email{Contact e-mail: guoliang@pmo.ac.cn (G. Li);}
\author[0000-0002-0478-3431]{Wenwen Zheng}
\affil{Purple Mountain Observatory, Chinese Academy of Sciences, Nanjing, Jiangsu, 210023, China}

\author[0000-0002-6506-1985]{Xiaoting Fu}
\affil{Purple Mountain Observatory, Chinese Academy of Sciences, Nanjing, Jiangsu, 210023, China}

\author[0000-0002-3759-1487]{Yang Chen}
\affil{School of Physics and Optoelectronic Engineering, Anhui University, Hefei, 230601, China}

\author[0000-0001-5284-8001]{Xuefei Chen}
\affil{Yunnan Observatories, Chinese Academy of Sciences (CAS), Kunming 650216, Yunnan, China}
\affil{School of Astronomy and Space Science, University of Chinese Academy of Sciences, Beijing, 100049, People's Republic of China}

\author[0000-0001-9989-9834]{Yanjun Guo}
\affil{Yunnan Observatories, Chinese Academy of Sciences (CAS), Kunming 650216, Yunnan, China}

\author[0000-0001-9387-7343]{Xuechun Chen}
\affil{Institute for Frontier in Astronomy and Astrophysics, Beijing Normal University, Beijing, 102206, China}
\affil{School of Physics and Astronomy, Beijing Normal University, Beijing 100875, China}

\author{Huanyuan Shan}
\affil{Shanghai Astronomical Observatory, Chinese Academy of Sciences, Shanghai, 200030, China}

\author{Guoliang Li}
\affil{Purple Mountain Observatory, Chinese Academy of Sciences, Nanjing, Jiangsu, 210023, China}

\begin{abstract}
Individual stars located near the caustics of galaxy clusters can undergo extreme magnification when crossing micro-caustics, rendering them observable even at cosmological distances.
Though most massive stars are likely reside in binary systems rather than as single star,  the influence of binary star system on magnification events is severely under-explored.
In this work, we simulate the light curves produced by detached binary stars crossing micro-caustics, aiming to characterize their unique observational signatures.  
Using high-resolution magnification maps generated by the GPU-PMO-CAUSTIC algorithm and \texttt{PARSEC} stellar models with red-shifted magnitude, we examined the impact of binary star parameters and crossing geometries on microlensing magnification patterns.
Our simulations reveal that binary stars produce diverse light curve features, including overlapping peaks, plateau-like structures, and time-variable color-magnitude differences. These features, particularly the distinct temporal variations in spectral energy distributions, offer diagnostic tools for distinguishing binary systems from single stars.
We further demonstrate the potential of multi-band photometry using the Chinese Space Station Telescope’s Multi-Channel Imager (CSST-MCI) to capture these variations. 
Our findings provide theoretical support for identifying binary systems in future caustic-crossing events, enabling more accurate characterization of high-redshift stellar populations.
\end{abstract}


\section{Introduction}
Galaxy clusters, the most massive gravitationally bound structures in the Universe, produce a diverse range of strong gravitational lensing effects. These include the formation of multiple images and giant arcs of background sources in photometric observations.
Acting as lenses, smoothly distributed matter surface density gives rise to smooth critical curves with extremely high magnifications in the image plane, corresponding to caustics in the source plane \citep{1992grle.book.....S}. 
Theoretically, the region near the critical curve, with its extremely high magnification, acts as a cosmic telescope, enabling the detection of individual stars at cosmological distances \citep{1991ApJ...379...94M}.
However, within strong lensing clusters, stars distributed in the intra-cluster light (ICL) and potentially existing Massive Compact Halo Objects (MACHOs) act as micro-lenses, disrupting the smooth macro- critical curve and producing a band of micro-critical curves, along with a corresponding micro-caustic band in the source plane \citep{2017ApJ...850...49V,2018ApJ...857...25D}. 
As a high-redshift star in the source plane moves across each micro-caustic, it experiences significant magnification, leading to a marked enhancement in brightness and resulting in transient brightness peaks.
When these peaks exceeds the instrument’s detection limit, they appear as transient sources, known as  ``caustic crossing events". 

These features give rise to transient magnification phenomena, as demonstrated by the detection of the first  high-redshift star ``Icarus'' \citep{2018NatAs...2..334K} and other candidates in recent studies. 
Transient sources at redshifts $z\sim$ 1-6 have been discovered along giant arcs crossing the macro critical curves of galaxy clusters, closely aligned with the critical curve itself \citep{2018NatAs...2..324R, 2019ApJ...881....8C,  2019ApJ...880...58K, 2022Natur.603..815W, 2022ApJ...940L..54C, 2022arXiv221102670K, 2023A&A...679A..31D, 2023ApJ...944L...6M, 2023MNRAS.521.5224M, 2023A&A...672A...3D, 2023ApJS..269...43Y, 2024arXiv240408045F}. Many of these sources are likely candidates for high-redshift stars undergoing caustic crossing events. 
With James Webb Space Telescope (JWST)\citep{2006SSRv..123..485G, 2023PASP..135d8001R}  and the upcoming Chinese Space Station Telescope (CSST)\citep{2023SCPMA..6619511C}, more extreme magnification events are expected in future observations \citep{2019A&A...625A..84D}. 
The magnification from these events provides a rare opportunity to study high-redshift stars, offering insight into stellar populations in the early universe \citep{2018ApJS..234...41W, 2024MNRAS.533.2727Z}, the role of MACHOs in dark matter halos \citep{2017ApJ...850...49V,2018ApJ...857...25D, 2018PhRvD..97b3518O, 2024arXiv240519422B, 2024arXiv240316989V}, and other dark matter substructures in galaxy clusters \citep{2018ApJ...867...24D, 2020AJ....159...49D, 2024ApJ...961..200W}.

It is widely discussed that most massive stars (greater than 15 $M_\odot$, here) are very likely to reside in binary systems rather than as single stars \citep{2012Sci...337..444S, 2014ApJS..215...15S, 2022A&A...667A..44G, 2024PrPNP.13404083C, 2024arXiv241118470B}. 
Observations of high-redshift star candidates suggest that some events cannot be fully explained by single-star models. For instance, the $z$=6.2 star Earendel, the $z$=2.091 star Mothra, and several other high-redshift star candidates show a double-peaked structure in their Spectra Energy Distribution (SED) or spectra, indicative of a multiple-star system \citep{2022ApJ...940L...1W, 2023A&A...679A..31D, 2024MNRAS.527L...7F, 2024arXiv240612607N}. 
Further insight comes from light curve analysis, as seen in the case of the Icarus event: the double peak in LS1’s light curve suggests it could correspond to a binary star system. 
Such findings underscore the necessity of understanding the unique light curve signatures of binary systems crossing micro-caustics, which may provide additional constraints on the nature of high-redshift sources.

In this work, we simulate the light curves produced by detached massive binary stars under varying conditions, such as orbital semi-major axes and crossing angles with the micro-caustics. By examining the effects of binary star parameters on light curves, we aim to establish diagnostic tools for distinguishing binary systems from single stars in caustic-crossing events.
High-resolution magnification maps of the microlensing source plane in extremely high-magnification regions were generated using the GPU-CAUSTIC code \citep{2023ChA&A..47..570Y}. To model binary star systems, we selected two pairs of single stars with similar ages from the \texttt{PARSEC} 1.2S stellar evolutionary models \citep{2012MNRAS.427..127B, 2015MNRAS.452.1068C}, calculating their intrinsic fluxes by convolving their redshifted spectra with filter transmission curves using the YBC code \citep{2019A&A...632A.105C}.
Assuming that the binary system’s center of mass moves at a constant velocity across the source plane, we modeled the orbital motion of each star around the center of mass as they crossed the micro-caustics, generating individual magnification curves. The total light curve was obtained by summing the magnified fluxes of both stars. 
Additionally, the CSST-MCI, capable of simultaneous photometric observations across ultraviolet to visible wavelengths, is designed for ultra-deep field surveys. Therefore, we simulated multi-band light curves and color-magnitude differences using MCI’s three broad filters to evaluate their effectiveness in distinguishing binary star systems.

The paper is organized as follows: Section~\ref{sec:methods} outlines the numerical methods employed in this study. 
Section~\ref{sec:single star cross a caustic} investigates whether the magnification curve of a uniformly luminous star crossing a micro-caustic aligns with the theoretical predictions for a star crossing a macro-caustic. 
Section~\ref{sec:Light curves of binary stars} presents the characteristic light curves of binary systems crossing micro-caustics, modeled under varying orbital parameters and geometries.
Section~\ref{sec:Lightcurves of binary stars from multi-color} presents multi-band light curves and color-magnitude differences for a binary star system crossing a micro-caustic, aiding in better distinguishing binary star systems. 
This section also highlights the potential of multi-band light curves obtained using the CSST-MCI, in identifying binary star systems.
A short summary is given in section~\ref{sec:Summary}.
Throughout this paper, we adopt the $\Lambda$CDM cosmological model with $h=0.7$ and $\Omega_m=0.308$.

\section{Methods} \label{sec:methods}
This section outlines the methods used in this study, including the simulation of microlensing magnification maps, the construction of binary star systems, and the introduction of the CSST-MCI instrument for multi-band observations. These components form the basis for analyzing the light curves of binary stars crossing micro-caustics.

\subsection{Simulate the microlensing magnification maps}
The first step in this study is to simulate microlensing magnification maps in the source plane. In these maps, the originally smooth macro-caustic is fragmented into a band of micro-caustics due to the influence of micro-lenses.
The commonly used method for generating such maps is the inverse ray-tracing algorithm \citep{1986A&A...166...36K, 1986MPARp.234.....S}. In this approach, a large number of rays are traced backward from the image plane, deflected by microlenses according to the lens equation, and projected onto a pixelated source plane. The magnification of each source plane pixel is determined by the number of rays that converge within it. However, this method imposes immense computational demands, especially for high-resolution maps, necessitating highly optimized algorithms \citep{1999JCoAM.109..353W, 2010NewA...15...16T, 2011ApJ...741...42M, 2021A&A...653A.121S}.

To address this, we developed the GPU-PMO code \citep{2022ApJ...931..114Z, 2023RAA....23h5011Z}, which employs a three-level grid structure combined with GPU parallel processing. This design significantly accelerates the ray-tracing procedure and improves computational efficiency. The GPU-PMO code is primarily used to generate magnification maps for regions of normal magnification, following the standard microlens equation.
However, the extremely high magnification regions near the macro-caustics, central to this work, require a modified micro-lens equation, as described in \citep{2017ApJ...850...49V}. Additionally, these regions demand even higher computational resources due to the extremely high magnification.
To tackle this challenge, we extended the GPU-PMO algorithm and developed the GPU-PMO-CAUSTIC code \citep{2023ChA&A..47..570Y}, specifically designed for simulating magnification maps in regions of extreme magnification near the macro-caustics. This enhanced algorithm retains the efficient structure of GPU-PMO while incorporating the necessary modifications to handle the unique computational demands of highly magnified regions.

To generate microlensing magnification maps, it is first necessary to model the strong lensing system of the galaxy cluster. In this work, we adopt the strong lensing parameters from \citep{2017ApJ...850...49V}, which were derived for the galaxy cluster MACS J1149 ($z_l = 0.54$), the host of the first high-redshift star ``Icarus" ($z_s \sim 1.5$)\citep{2018NatAs...2..334K}.
For micro-lenses, the magnification maps in this study adopt two different surface densities: $\kappa_* = 0.0004$ and $\kappa_* = 0.002$, corresponding to $\Sigma_* = 1 M_\odot/\mathrm{pc}^2$ and $5 M_\odot/\mathrm{pc}^2$, respectively. 
These values are lower than those used in \cite{2017ApJ...850...49V}. Since this study primarily focuses on the light curves of binary stars crossing individual micro-caustics, a lower $\kappa_*$ is chosen to ensure minimal overlap of the micro-caustics. 
The micro-lenses are uniformly distributed with an average mass of $\langle M \rangle = 0.3 M_\odot$.
Given a lens redshift of $z_l = 0.54$ and a source redshift of $z_s = 1.5$, the corresponding microlensing Einstein radius $\theta_E=\sqrt{4G\left\langle M\right\rangle D_{ls}/({c^2}{D_lD_s})} \sim 10^{-6}$ arcsec, which also serves as the length unit for the magnification map.

\subsection{The broad band filters for CSST-MCI}
The Multi-channel Imager (MCI) of the Chinese Space Station Telescope (CSST) is a specialized imaging instrument capable of simultaneous tri-channel exposures. It enables photometric observations across three broad bands, covering the full wavelength range from ultraviolet to visible light \citep{2022RAA....22b5019C}.
The MCI will be installed at the focal plane of the 2-meter main optical system of the multifunctional optical facility aboard the space station. One of its primary scientific objectives is the observation of ultra-deep fields in ultraviolet and optical wavelengths, including two blank deep fields and four galaxy cluster deep fields. It is anticipated to achieve the deepest ultraviolet and optical fields ever observed, enabling studies such as galaxy cluster gravitational lensing, Type Ia supernova cosmology, the joint evolution of galaxies and black holes at the low-mass end, and the detection of extremely magnified stars through gravitational lensing.

The CSST-MCI offers significant advantages in the discovery and study of extremely magnified stars through gravitational lensing. First, the MCI’s wide field of view, approximately 60 square arcminutes, facilitates the detection of caustic-crossing events. Additionally, its capability for simultaneous tri-color observations ensures the highest possible accuracy in photometric spectral energy distributions (SEDs). Furthermore, the MCI allows for exposures of up to 1500 seconds on a target. With its ultra-wide filters, repeated exposures can be stacked to achieve a depth of up to 30 magnitudes, making it ideal for follow-up monitoring of light curves of candidate high-redshift stars undergoing extreme magnification.

In this work, we simulate light curves using MCI’s three ultra-wide filters CBU, CBV, and CBI. The intrinsic fluxes of stars, prior to lensing magnification, are determined by their fluxes transmitted through these filters.

\subsection{Construct binary systems}

\begin{table*}
\centering
\caption{Model Parameters for Binary Star System 1 (Star A \& Star B) and System 2 ($\text{Star A}^{\prime}$ \& $\text{Star B}^{\prime}$). The two binary systems are placed at a redshift of 
z=1.5, and the Intrinsic AB magnitude refers to the flux of the stars at this redshift before
being magnified by lensing.}
\begin{tabular}{llllll}
\hline
Binary Parameter & Symbol & star A  & star B  & $\text{star A}^{\prime}$  &  $\text{star B}^{\prime}$\\
\hline
Initial stellar mass  & $M_I\,(M_\odot)$ & $90$  &$80$  &$70$ &$60$\\
Current stellar mass  & $M\,(M_\odot)$ & $87.26$  &$78.10$  &$68.46$ &$58.98$\\
Stellar radius        & $R\,(R_\odot)$ & $140$ &$45$  &$40$  &$20$ \\
Bolometric luminosity & $L\,(L_\odot)$ & $1.89\times10^{6}$ & $1.51\times10^{6}$ & $1.24\times10^{6}$  &$9.01\times10^{5}$\\
Temperature & $T_{eff}\,(K)$ & $18127$ & $30315$ &$30356$ &$38298$\\
Age & $Age (Myr)$ & $3.151$ & $3.152$  &$3.401$  &$3.401$\\
Intrinsic AB magnitude in CBU band & $m_\mathrm{CBU}$ & $34.80179$  &$34.88679$  &$35.10233$  &$35.76134$\\
Intrinsic AB magnitude in CBV band & $m_\mathrm{CBV}$ & $34.63210$  &$35.21515$  &$35.43049$  &$36.23490$\\
Intrinsic AB magnitude in CBI band & $m_\mathrm{CBI}$ & $34.74987$  &$35.74113$  &$35.95656$  &$36.83487$\\
\hline
\end{tabular}
\label{tab:table1}
\end{table*}

\begin{figure}
	\includegraphics[width=\linewidth]{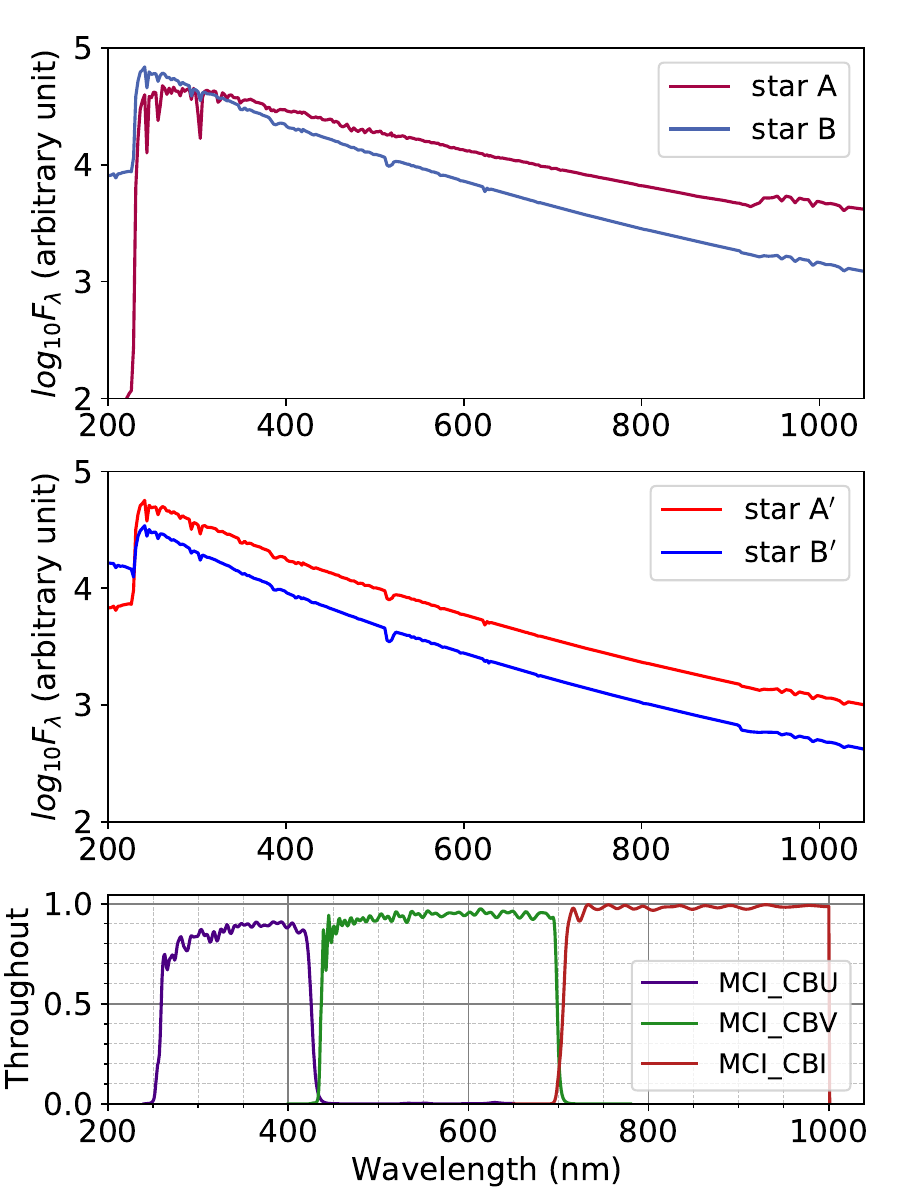}
    \caption{The spectra of the for stars (redshited to $z =1.5$) and the throughout of the three broad bands of the CSST-MCI. 
    }
    \label{fig:star_spectra_filters}
\end{figure}

\begin{figure*}
	\includegraphics[width=\textwidth]{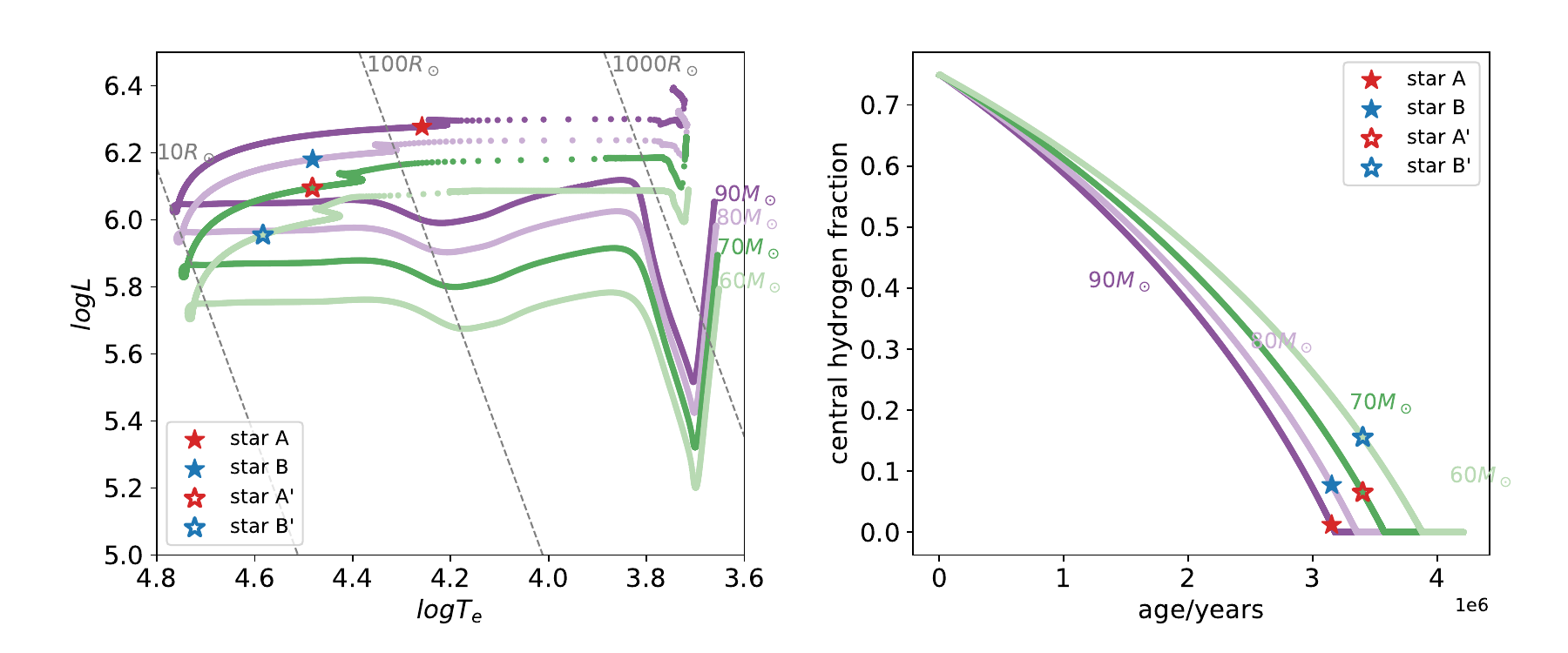}
    \caption{ H-R diagram and central hydrogen fraction vs. age relation for stars with masses of $90M_\odot$, $80M_\odot$, $70M_\odot$, and $60M_\odot$. Star A and Star B, which form binary system 1, are marked with solid pentagrams, while $\text{Star A}^{\prime}$ and $\text{Star B}^{\prime}$, which form binary system 2, are marked with hollow pentagrams.
    }
    \label{fig:HR_age_HCEN}
\end{figure*}

In this study, we focus on detached binary systems, which are modeled using the single star evolutionary tracks from \texttt{PARSEC} v1.2S \footnote{\url{http://stev.oapd.inaf.it/PARSEC/tracks_v12s.html}} \citep{2012MNRAS.427..127B, 2015MNRAS.452.1068C}. Each star in the binary system is treated as an independent entity and orbit with each other. Mass-loss induced by stellar winds has a large impact on magnitude and color of stars, as well as the current stellar mass which affect the binary motion. 
The \texttt{PARSEC} models we use adopt sophisticated mass-loss recipes for massive stars and can reproduce the observations of local massive stars.

Stars with the same age and composition (formed together) are paired to form a binary.
In this work, we focus on bright high-mass stars ($50 M_\odot < M_I < 100 M_\odot$) 
as binary components. 
We choose low metallicity ($Z = 0.001$) to match the stellar population at the  redshift $z_s =1.5$ of interest. 
Two pairs of binary system with four stars are selected.  
Table~\ref{tab:table1} summarizes the parameters of the two systems, which include stellar masses, radii, luminosities, temperature, age and the intrinsic magnitude at $z_s =1.5$  in the CSST-MCI photometry system. 

The photometry and flux of these stars is calculated using the method of the  \texttt{YBC}  stellar bolometric correction database \footnote{\url{https://sec.center/YBC/}} \citep{2019ApJ...881....8C}.
Unlike local universe stars, applying bolometric corrections to high-redshift stars necessitates accounting for the red-shifted flux within a specific photometric system.  
Currently, no publicly available database considers redshift in bolometric corrections.
To address this gap, we are going to release a new version of \texttt{YBC} database (Chen, Fu, et al., in prep.). This work presents the first application of the \texttt{YBC} database with redshift.

For the two pairs of selected binary systems,
system 1 consists of Star A and Star B, with the initial mass of $90 M_\odot$ and $80 M_\odot$, respectively, while System 2 comprises $\text{Star A}^{\prime}$ and $\text{Star B}^{\prime}$, with initial masses of $60 M_\odot$ and $50 M_\odot$. 
Since massive stars lose mass via stellar wind during their evolution, we use their current masses to calculate the orbital dynamics of the binary systems. 
From the H-R diagram and the central hydrogen fraction vs. age relation shown in Fig.~\ref{fig:HR_age_HCEN}, the stars in both systems are near the end of the main sequence, with Star A in System 1 transitioning toward the blue supergiant (BSG) phase.
The spectra of these four stars (redshited to $z =1.5$) are shown in Fig.~\ref{fig:star_spectra_filters}.
Regarding the choice of orbital parameters for the binary systems, we defer this to Section~\ref{sec:Light curves of binary stars}.

In this first approach of characterizing binary stars in lensing events, additional complexities such as  stellar deformation, dust obscuration, and mass transfer are not considered. We plan to incorporate these in future work to explore their effects on light curve characteristics.

\section{The Magnification of a single star cross a micro caustic} \label{sec:single star cross a caustic}
Before exploring the light curves of binary stars crossing micro-caustics, we first aim to understand the characteristic light curve profile produced by a single star crossing a single micro-caustic. Since the light curve of a single star is entirely determined by its magnification, this section focuses on the magnification curve of a single star crossing a micro-caustic, without considering any specific stellar model.

\subsection{Theoretical magnification curve} \label{subsec: Theoretical}

The magnification curve of a star passing through the macro-caustic of a galaxy cluster was first explored by \citep{1991ApJ...379...94M}, who first demonstrated that the extreme magnification could render the star detectable even at cosmological distances.
In their model, the star is treated as a circular disk with an angular radius $R$ and  uniform surface brightness, located at a distance $l$ from the caustic. 
The total magnification $\mu$ is obtained by integrating the point-source magnification across  the star's surface, expressed as
\begin{equation}
\mu=\frac{\mu_0}{R^{1/2}} F\left ( \frac{l}{R}-1  \right ),
\end{equation}
where
\begin{equation}
\begin{aligned}
    &F(y_0)=\frac{2}{\pi}\int_{-y_0}^{2} \mathrm{d}y\left [ \frac{y(2-y)}{y+y_0}  \right ]^{1/2}  (y_0<0), \\
    &F(y_0)=\frac{2}{\pi}\int_{0}^{2} \mathrm{d}y\left [ \frac{y(2-y)}{y+y_0}  \right ]^{1/2}  (y_0>0),
\end{aligned}
\label{eq:theoratical_curve}
\end{equation}
and $\mu_0$ is a parameter related to the strong lens model of the cluster, as described in Eq.(5) from \cite{1991ApJ...379...94M}.

\begin{figure}
	\centering
	\includegraphics[width=\linewidth]{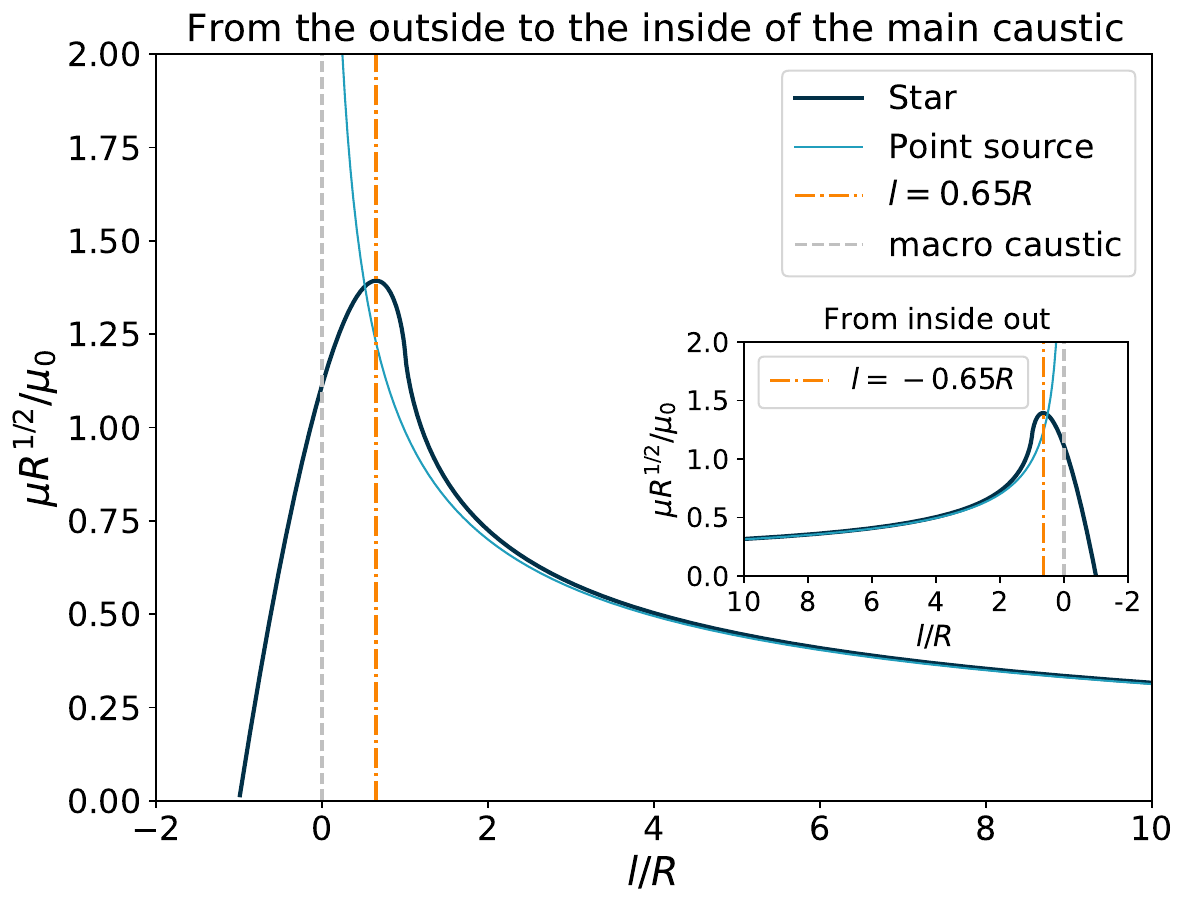}
    \caption{Theoretical magnification curve of a uniformly luminous star as a function of its position relative to the macro-caustic.
    The vertical axis represents the magnification $\mu$ multiplied by the factor $R^{1/2}/\mu_0$, where R denotes to the star radius and $\mu_0$ is a parameter related to cluster strong lens model.
    The dark blue and light blue line denotes the magnification of uniformly luminous star and a point source, respectively. The gray dashed line and orange dash dotted line denotes the position of the macro-caustic and the peak position of star magnification, respectively. 
    }
    \label{fig:Theoretical_magnification_curve}
\end{figure}

\begin{figure*}	
	\centering
	\includegraphics[width=\linewidth]{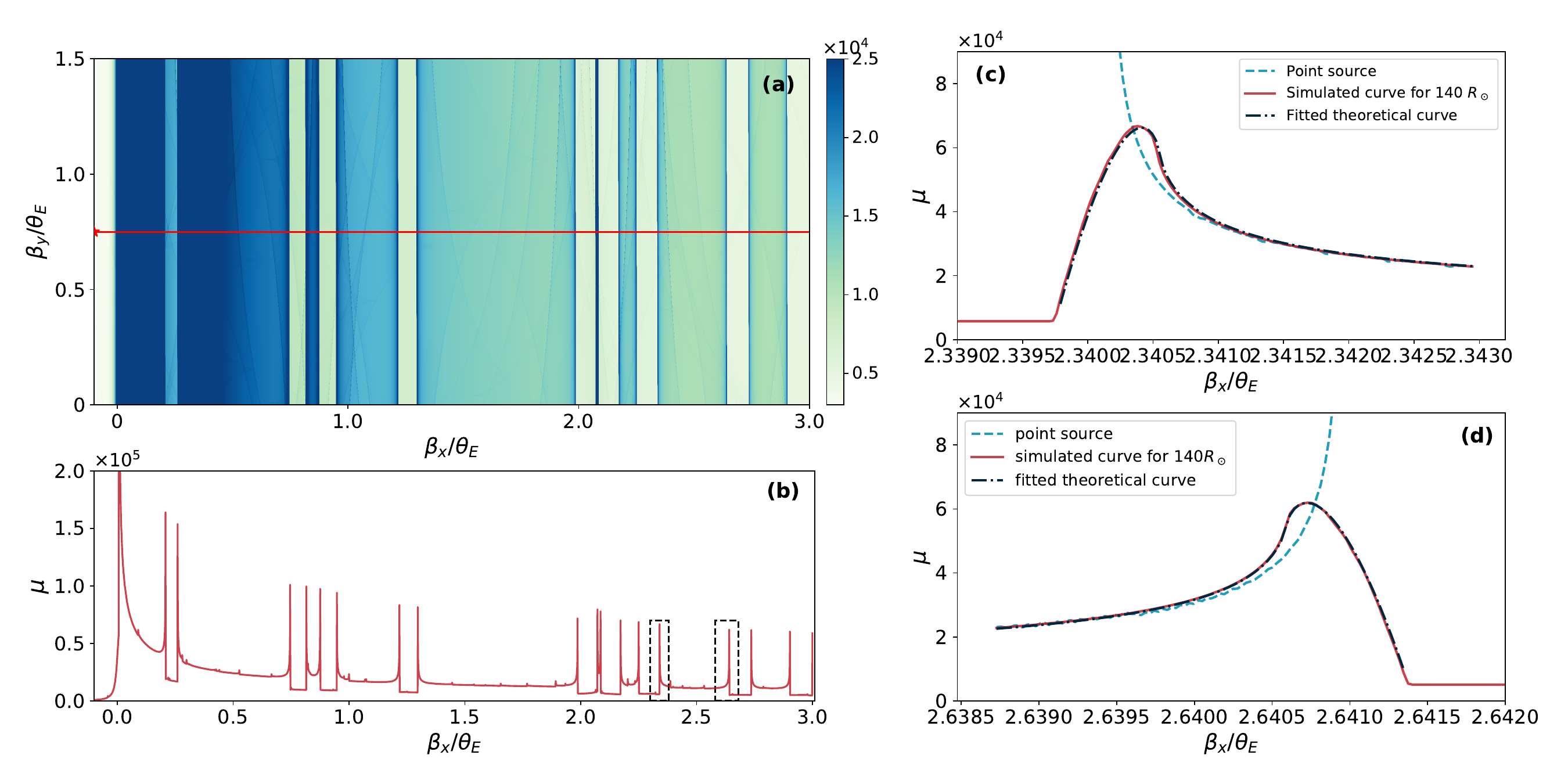}
    \caption{
    Microlensing magnification map and the magnification curve of a single star with uniform surface brightness. 
    Panel (a) displays a partial magnification map with a resolution down to 10 $R_\odot$, the colorbar represents the magnitude of the magnification.
    The red solid line represents a trajectory entering the micro caustic band from the outside. 
    Panel (b) presents the magnification curve produced by a star with a radius of 140 $R_\odot$ moving along this trajectory. 
    Panels (c) and (d) present zoomed-in views of two selected peaks (highlighted by black dashed boxes in panel (b)), showing the alignment between the simulated magnification curve (solid red) and the shifted theoretical curve (black dashed).
    }
    \label{fig:Simulation_and_fitting_of_mianification_curve}
\end{figure*}

Following these equations, Fig.~\ref{fig:Theoretical_magnification_curve} illustrates the magnification as the source star moves perpendicularly from outside to inside the macro-caustic and vice versa. 
Here, the x axis has been converted from representing time to denoting the relative position of the star with respect to the caustic.
These curves exhibit certain distinct characteristics, as also noted in \citep{1991ApJ...379...94M}. The key feature of this theoretical curve is that the peak magnification does not occur when the star’s center aligns exactly with the caustic ($l= 0$). Instead, it reaches a maximum at  $l= 0.65R$.
This is likely due to the integration of magnification across the uniformly luminous circular source. 

This characteristic profile is considered a diagnostic signature of single stars crossing macro-caustic. 
However, due to the presence of microlenses in galaxy clusters, the smooth macro-caustic is replaced by a band of micro-caustics. In the following, we investigate whether the magnification variation curve for a single star crossing an individual micro-caustic within this band still closely follows the theoretical curve.

\subsection{Numerical simulation and fitting} \label{subsec: simulation}

Due to the complexity of magnification distributions introduced by microlensing effects, deriving the magnification variation analytically becomes challenging. In this study, we employ numerical simulations to assess whether the magnification curve for a single star crossing an individual micro-caustic aligns with the theoretical predictions. 
Specifically, we simulate a large number of magnification curves for single stars crossing micro-caustics and fit them to the theoretical model.

We first generate microlensing magnification maps, with details provided in Section~\ref{sec:methods}. One partial magnification map used in this analysis is shown in Fig.~\ref{fig:Simulation_and_fitting_of_mianification_curve} panel (a), featuring a spatial resolution of approximately 10$R_\odot$. The spatial coordinates are expressed in units of $\theta_E$, with $\beta_x=0$ marking the original position of the macro-caustic.
In the magnification map, the color shading represents the magnification levels, with the vertical and slanted line structures indicating the micro-caustics. 
To facilitate subsequent peak fitting, we adopt a low micro-lens surface density of $\kappa_*=0.0004$ ($\Sigma_*= 1 M_\odot/\mathrm{pc}^2$). This sparse distribution reduces overlap between micro-caustics. Additionally, a lower microlens surface density leads to higher magnifications for individual micro-caustics, which is reflected in the resulting magnification curves.
 
A uniformly luminous star of radius 140 $R_\odot$ moves along a straight trajectory (marked in red in panel (a) of Fig.~\ref{fig:Simulation_and_fitting_of_mianification_curve}) from left to right across the magnification map, generating a corresponding magnification curve shown in panel (b). The magnification at each point is calculated by averaging the magnifications of all pixels covered by the circular source area as its center progresses along the trajectory.
As the star traverses this path, it crosses multiple micro-caustics, resulting in a magnification curve characterized by a series of distinct peaks, each corresponding to an individual micro-caustic crossing.
We extract the peaks from the magnification curve shown in panel (b), with each peak corresponding to the star crossing  micro-caustics.

To assess the alignment between these simulated peaks and theoretical curves,
we fit the theoretical curve $f(x)$ (as $F(y_0)$ described in eq.(\ref{eq:theoratical_curve}) and is represented by the dark blue line in Fig.\ref{fig:Theoretical_magnification_curve}),
to the simulated magnification curve $S(x)$ (solid red line) by shifting $f(x)$.
The fitting process involved applying positional offsets (C) and scaling factors (A and B) to account for the location of the micro-caustic and the lensing environment’s overall magnification:
\begin{equation}
    S(x) = A+Bf(x+C).
\end{equation}
Specifically, $C$ represents the position of the micro-caustic and can be determined from the position of the peak magnification of a point source,
$B$ reflects the magnification contribution from strong lensing, while $A$ accounts for additional magnification effects introduced by the microlensing environment. Both $A$ and $B$ are primarily obtained through a $\chi^2$ minimization method.
The panel (c) and (d) of Fig.~\ref{fig:Simulation_and_fitting_of_mianification_curve} highlights two selected peaks (highlighted with black dashed boxes in panel (c)) from the simulated magnification curve, showing excellent agreement with the shifted theoretical model. 

Based on the fits across numerous peak values, we conclude that when the density of micro-caustics is low and overlap is minimal, most peak values align closely with the theoretical model, suggesting that the magnification characteristics of a uniformly luminous star crossing a single micro-caustic largely follow theoretical predictions. 
This consistency generally holds except when micro-caustics are closely spaced. In such cases, the magnification curve exhibits a more complex structure influenced by the source's size and the distribution of the micro-caustics.
However, our fitting approach here is primarily focused on scenarios where the source crosses a single micro-caustic, which allows us to better examine the alignment between simulated and theoretical curves.

Exploring the magnification curves of single stars crossing individual micro-caustics forms the foundation for interpreting observed light curves and discerning stellar properties in high-magnification lensing events. These curves provide a crucial baseline for analyzing more complex scenarios, such as binary star systems, which are explored in subsequent sections.


\section{Binary stars cross a caustic} \label{sec:Light curves of binary stars}
Having characterized the magnification curve for a single star crossing a single micro-caustic, we now extend our analysis to binary star systems.
Binary star systems present a more complex scenario for gravitational lensing than single stars due to their orbital motion. As binary stars orbit their center of mass, their combined motion relative to the observer and the micro-caustic generates unique magnification and light curves. Understanding these patterns provides the base knowledge of identifying binary systems in caustic-crossing events and distinguishing them from single-star systems in high-redshift observations.

To explore these features, we simulate the magnification and light curves of two detached massive binary star systems crossing a micro-caustic. The physical parameters of the binary stars are provided in Section~\ref{sec:methods}. However, binary star orbits can encompass highly complex parameter combinations, with intricate motion patterns reflected in the morphology of their light curves.
In this study, we adopt a simplified parameter model with a face-on circular orbit, treating the semi-major axis $Axis$ (equivalent to the orbital radius in the case of a circular orbit) as a free parameter.

As mentioned earlier, we place the binary star systems at a redshift of 1.5 and assume that they remain unresolved by observational instruments. Consequently, the observed flux is the sum of the individual fluxes of the two stars, each undergoing microlensing magnification as they traverse the micro-caustics. For this analysis, we use the flux in the CBU filter as the intrinsic flux prior to lensing, as an illustrative example.
In the source plane, the binary system exhibits orbital motion around its center of mass, while the combined relative motion of the observer, lens, and source causes the center of mass to traverse the source plane. As a result, the binary stars orbit each other while simultaneously crossing the micro-caustics.
We assume the center of mass of the binary system moves at a velocity of approximately $v = 800$ km/s \citep{2024ApJ...964..160H}. The light curve simulations cover a period of 40 days in the source rest frame, corresponding to about 100 days in the observer’s frame after accounting for the redshift factor $(1 + z)$.
To better resolve small-sized stars, we employ magnification maps with a resolution of approximately $5 R_\odot$, and the micro-lens surface density is set to $\kappa_*= 0.002$ ($\Sigma_* = 5 M_\odot/\mathrm{pc}^2$).

\begin{figure*}
	\centering
	\includegraphics[width=0.9\linewidth]{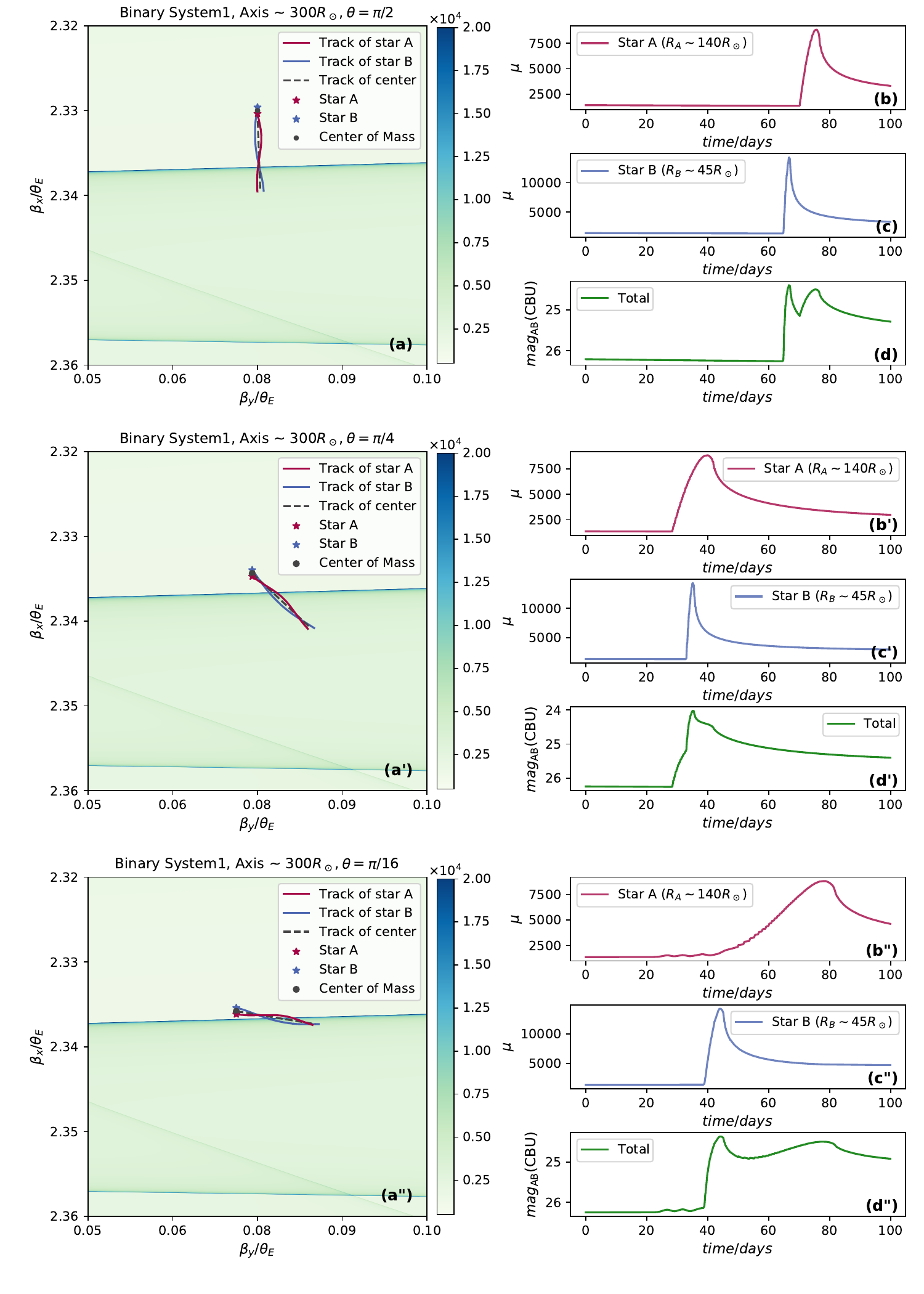}
    \caption{
    The motion trajectory of binary star system 1, consisting of Star A and Star B, is displayed on the zoomed-in magnification map, along with the corresponding magnification and light curves. The colorbar in the map represents the magnification values.
    In this figure, the binary system's orbital semi-major axis is fixed at 300 $R_\odot$ ($P \sim 50$ days). The trajectories and light curves are presented for three crossing angles: $\theta = \pi/2$, $\pi/4$, and $\pi/16$. Each set includes the binary stars' trajectories on the magnification map, the individual magnification curves for each star (red and blue curves), and the total light curve of the system (green curve).
    The detailed parameters of Star A and Star B are provided in Table~\ref{tab:table1}.
    }
    \label{fig:Binary_lightcurves_axis300}
\end{figure*}

\begin{figure*}
	\centering
	\includegraphics[width=0.9\linewidth]{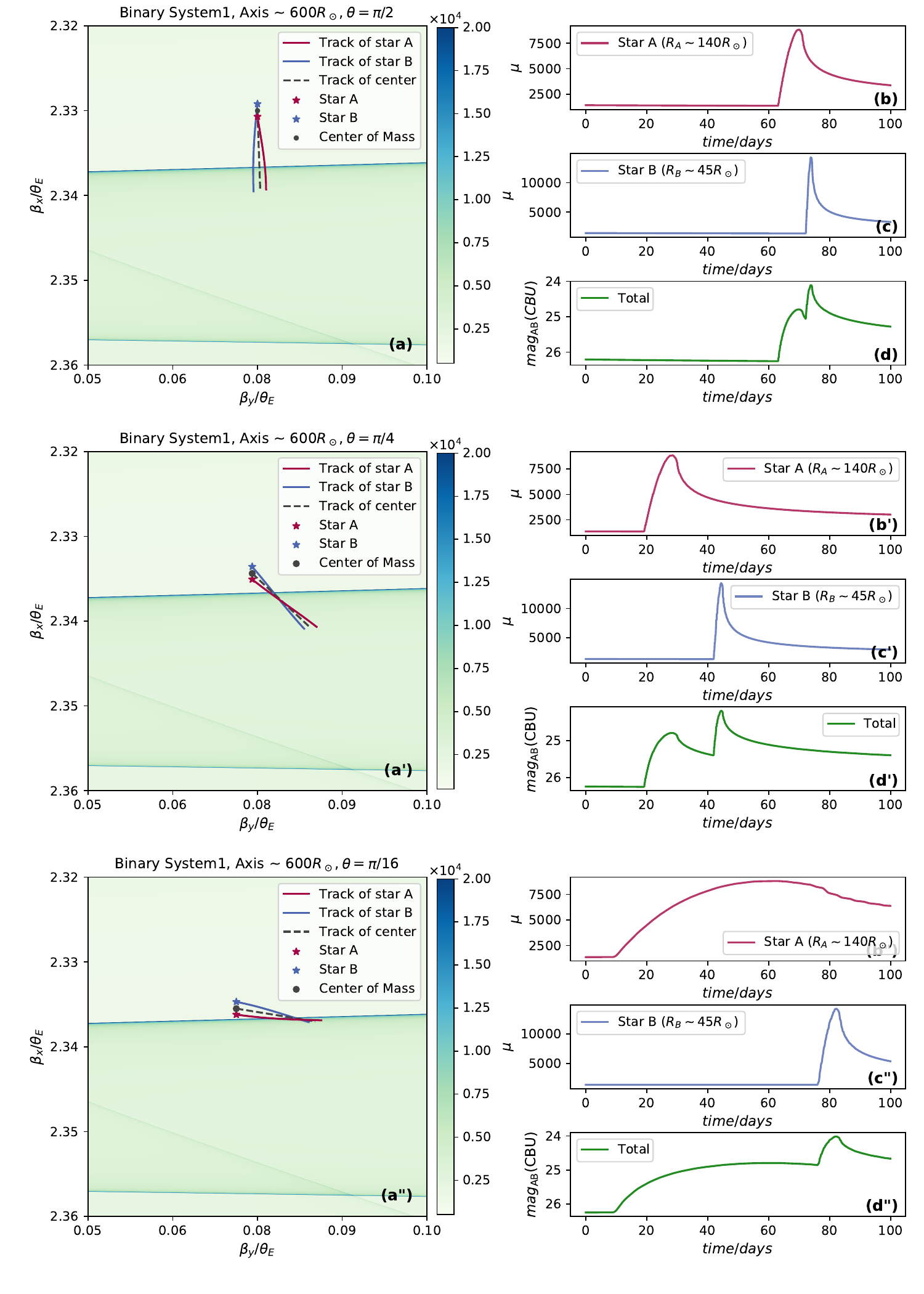}
    \caption{The motion trajectory, zoom-in magnification map, and corresponding magnification and light curves for the binary star system 1 (Star A and Star B) with an orbital semi-major axis of 600 $R_\odot$ ($P \sim 140$ days). 
    }
    \label{fig:Binary_lightcurves_axis600}
\end{figure*}

\begin{figure*}
	\centering
	\includegraphics[width=0.9\linewidth]{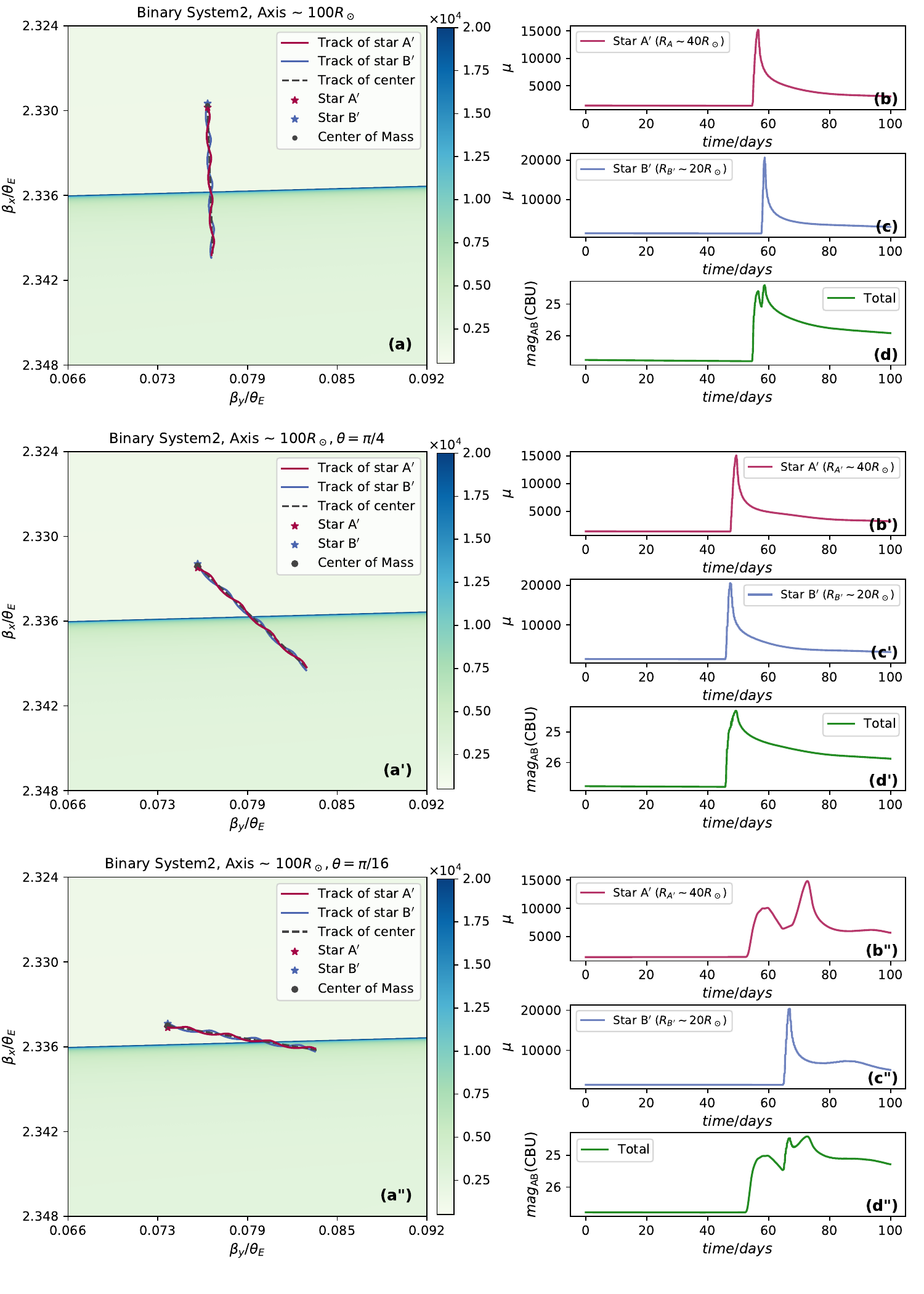}
    \caption{The motion trajectory, zoom-in magnification map, and corresponding magnification and light curves for the binary star system 2 ($\text{star A}^{\prime}$ and $\text{star B}^{\prime}$) with an orbital semi-major axis of 100 $R_\odot$ ($P \sim 11$ days).  
    The detailed parameters of Star A and Star B are provided in Table~\ref{tab:table1}.
    }
    \label{fig:Binary_lightcurves_axis100}
\end{figure*}

Fig.~\ref{fig:Binary_lightcurves_axis300} to Fig.~\ref{fig:Binary_lightcurves_axis100} illustrate the trajectories magnification maps, binary star trajectories, and resulting light curves. Different cases of the orbital radii is applied,  $Axis = 300 R_\odot$ (orbital period $P \sim 50$ days, see Fig.~\ref{fig:Binary_lightcurves_axis300}) and $Axis = 600 R_\odot$ ($P \sim 140$ days, see Fig.~\ref{fig:Binary_lightcurves_axis600}) are applied to the System 1 binary system,  while  $Axis = 100 R_\odot$ ($P \sim 11$ days, see Fig.~\ref{fig:Binary_lightcurves_axis100}) is applied to the System 2 binary system.
Each figure contains three panels, illustrating cases with different angles between the center-of-mass trajectory and the micro-caustics:  $\pi/2$, $\pi/4$
and $\pi/16$ in the top, middle, and bottom panels, respectively.
Additionally, the caustic in the figure forms an angle of approximately $\pi/90$ with the horizontal direction, which was included in our calculations.
Furthermore, in this work, we do not consider the sampling of the light curves; thus, the magnification and total light curves shown in the figures can be regarded as idealized curves.

We observe that, with the center-of-mass velocity held constant, the binary star masses, orbital radius $Axis$, and angle $\theta$ can collectively generate a diverse range of magnification patterns and light curve shapes.
Several key patterns emerge:

\begin{itemize}
\setlength{\itemindent}{0.8em}
\setlength{\leftmargin}{0.5em}
\setlength{\itemsep}{1.0ex} 
\setlength{\labelsep}{0.5em} 
\setlength{\listparindent}{0.8em} 
    \item \pmb{Effect of angle $\theta$} \\
           Smaller angles ($\theta= \pi/16$) lead to elongated light curves, as the stars spend more time in the high-magnification region. Larger angles ($\theta= \pi/2$) produce narrower, more distinct peaks.
           Theoretically, the angle $\theta$ between the center-of-mass trajectory and the micro-caustic significantly influences the duration of the stars' interaction with the caustic, which directly affects the width of the peaks in the magnification curve.
           As shown in three Figures, decreasing $\theta$ broaden the peak. When $\theta = \pi/16$ and the system is very close to the micro-caustic, at least one star in the binary system moves nearly parallel to the caustic, remaining in the high-magnification region for a longer time. Meanwhile, the other star is also magnified as it crosses the caustic, resulting in a broad double peak in the total light curve or a single peak superimposed on a hump. This indicates that the total light curve maintains a relatively high brightness with smaller variations over time.

    \item \pmb{Effect of orbital radius $Axis$} \\
           Larger orbital radii produce more isolated and distinct magnification peaks, as the stars traverse separate regions of the micro-caustic. Conversely, smaller radii result in overlapping peaks, as the stars cross the caustic in rapid succession.
           The orbital radius $Axis$ governs the spatial separation and rotational dynamics of the binary stars as they cross the caustic. As $Axis$ increases, the stars move along less curved, more extended paths. When $Axis$ becomes sufficiently large, their trajectories through the caustic approach straight lines.
           This leads to simpler, more isolated peaks in the light curve. Conversely, when $Axis$ is smaller, the orbit becomes more compact, increasing the curvature of each star’s path. 
           This can result in repeated caustic crossings, producing complex and closely spaced peaks in the light curve, particularly for lower mass binary systems, as shown in Fig.~\ref{fig:Binary_lightcurves_axis100}. 
           In principle, such systems can have smaller $Axis$. When $\theta$ is also small (bottom panel of Fig.~\ref{fig:Binary_lightcurves_axis100}), one of the stars in the binary system can even cross the caustic twice, leading to a light curve where a double peak overlaps with a single peak.

    \item \pmb{Effect of crossing sequence and interval} \\
           The sequence and timing of each star’s caustic crossing further influence the shape of the light curve.
           In the top panels of Fig.~\ref{fig:Binary_lightcurves_axis300}, Fig.~\ref{fig:Binary_lightcurves_axis100} and middle panel in Fig.~\ref{fig:Binary_lightcurves_axis600}, the two stars cross the caustic in sequence with noticeable intervals. 
           This produces two separate peaks in the total light curve, spaced closely in time but distinctly visible.
           However, in the middle panels of of Fig.~\ref{fig:Binary_lightcurves_axis300}, Fig.~\ref{fig:Binary_lightcurves_axis100} and top panel of Fig.~\ref{fig:Binary_lightcurves_axis600}, both stars cross the caustic nearly simultaneous, creating overlapping peaks in the light curve. 
           In some cases, as seen in the middle panel of Fig.~\ref{fig:Binary_lightcurves_axis100}, the total light curve from both stars may even resemble a single peak.
           
    \item \pmb{Effect of binary mass} \\
           For relatively lower-mass binary systems (System 2 in Fig.\ref{fig:Binary_lightcurves_axis100}), as mentioned earlier, they can have smaller orbital semi-major axes, allowing for multiple caustic crossings. Additionally, stars with smaller masses may exhibit narrower peaks due to their smaller radii, as shown in the top panel of Fig.\ref{fig:Binary_lightcurves_axis100}, which places higher demands on observational sampling. 
           Notably, a more pronounced mass ratio between the binary stars could further complicate the scenarios described above.
\end{itemize}

In future observations of caustic-crossing events, if the observed light curves exhibit the above characteristics, they could potentially aid in identifying the source as a binary star system.
However, actual observational scenarios are expected to be more complex, incorporating various observational effects and strategies, as well as additional complications from binary-specific effects such as stellar deformation, dust obscuration, orbital inclination, or mass exchange between massive binary stars.

\section{Light curves of binary stars from multi-color instrument of CSST  } \label{sec:Lightcurves of binary stars from multi-color}
To further explore the potential for identifying binary star systems, we conducted simulations of multi-band light curves for a binary star crossing a micro-caustic. 

Multi-band photometry across different wavelength offers a powerful tool for characterizing binary star systems caustic-crossing events. While gravitational lensing is achromatic, the distinct magnification levels experienced by each of the binary components as they cross a micro-caustic at different times lead to temporal variations in the observed SED. These variations manifest as changes in color-magnitude differences between photometric bands, providing a diagnostic feature for binary systems.

\begin{figure*}
	\centering
	\includegraphics[width=\linewidth]{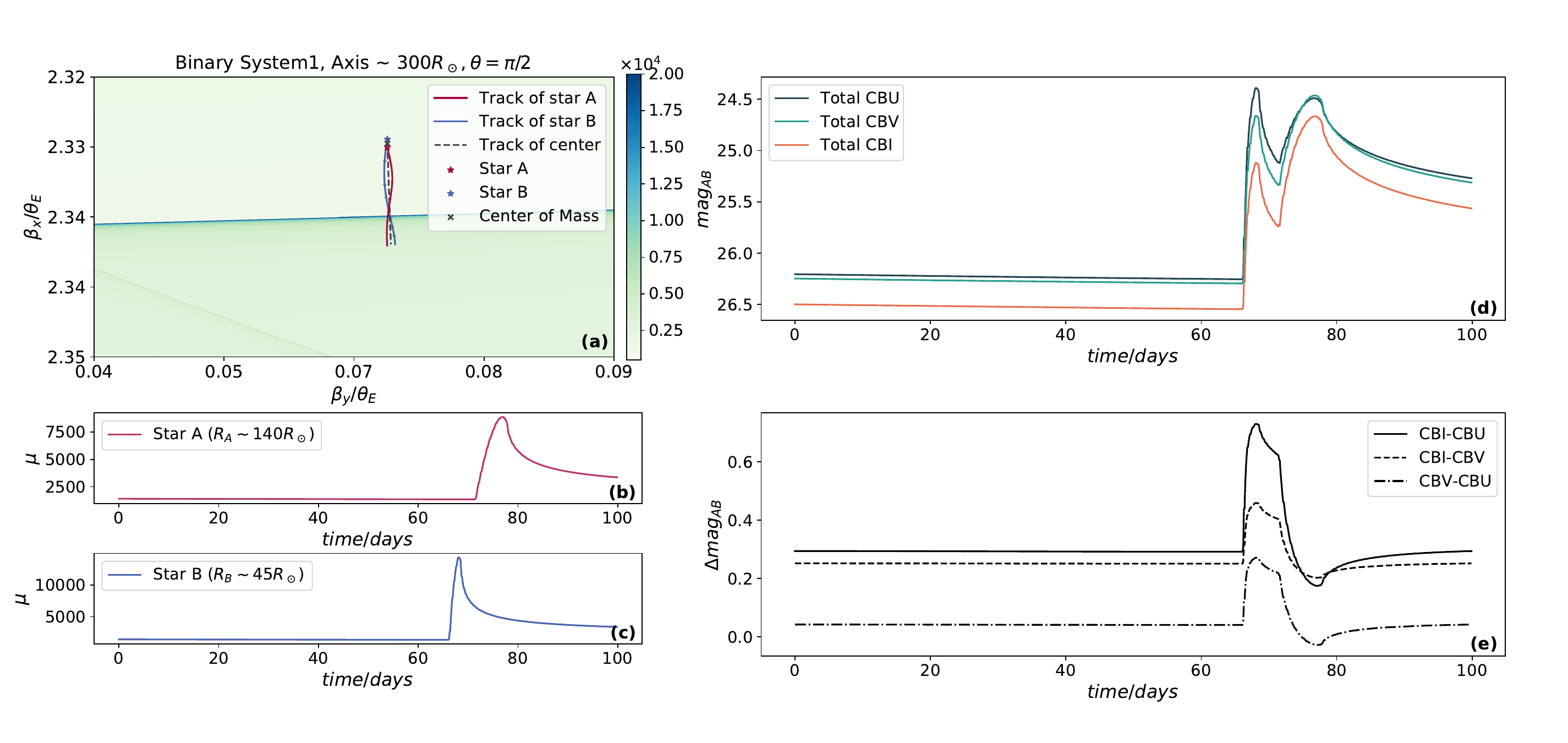}
    \caption{The simulated magnification curves, multi-band light curves, and color-magnitude differences for binary star system 1 crossing a micro-caustic at a crossing angle of $\theta = \pi/2$, using the three ultra-broad filters (CBU, CBV, CBI) of the CSST-MCI.
    Panel (a) shows the zoomed-in magnification map with the motion trajectories of the binary stars. Panels (b) and (c) display the magnification curves for Star A and Star B, respectively. Panel (d) presents the total light curves of  binary system 1 in the CBU, CBV, and CBI filters. Panel (e) illustrates the color-magnitude differences between the filters.
    }
    \label{fig:Binary_lightcurves_3filters_M90M80}
\end{figure*}

\begin{figure*}
	\centering
	\includegraphics[width=\linewidth]{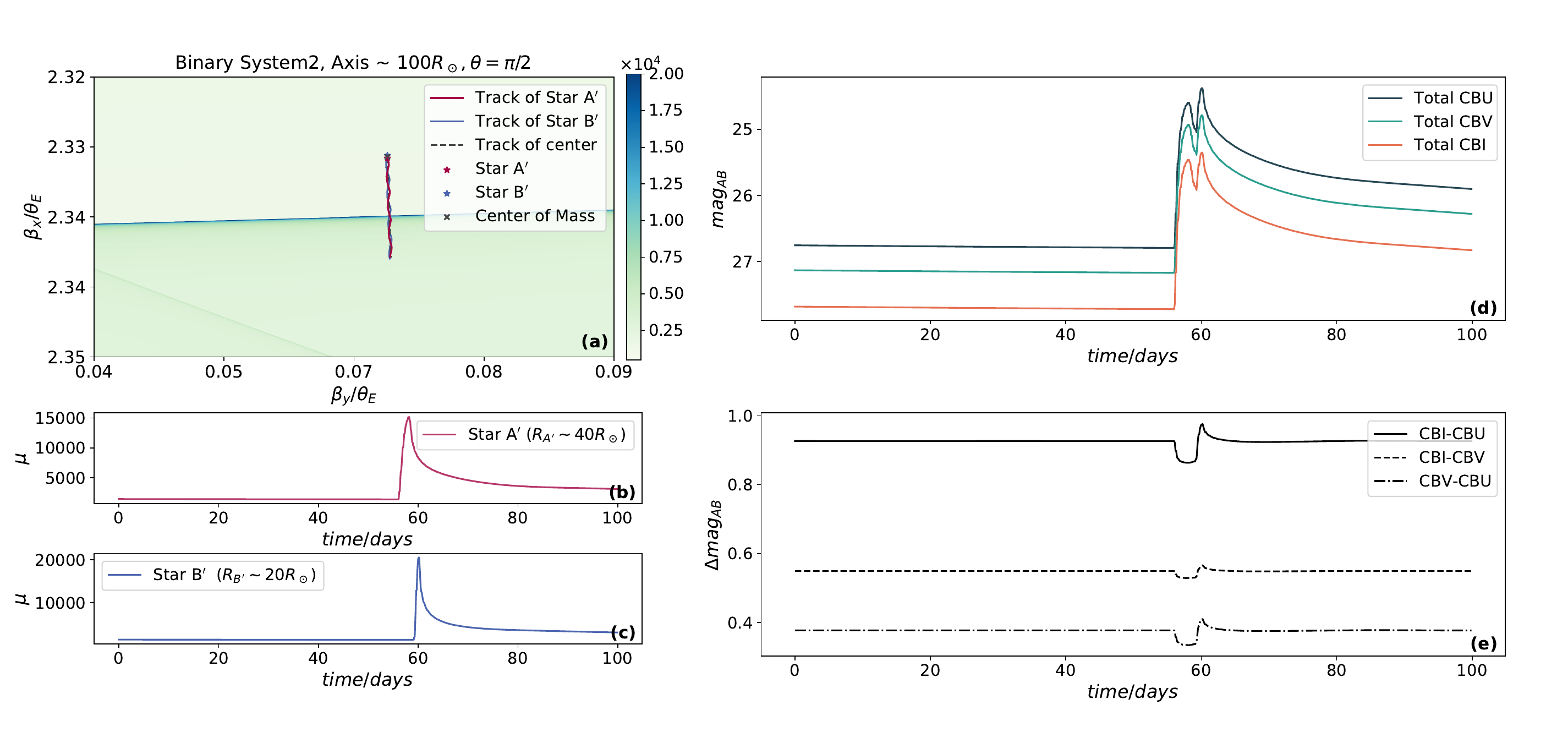}
    \caption{The simulated magnification curves, multi-band light curves, and color-magnitude differences for binary star system 2 crossing a micro-caustic at a crossing angle of $\theta = \pi/2$, using the three ultra-broad filters (CBU, CBV, CBI) of the CSST-MCI.
    }
    \label{fig:Binary_lightcurves_3filters_M70M60}
\end{figure*}

To clarify, for binary stars orbiting their center of mass and sequentially crossing the same caustic, each component produces its own peak in the light curve. 
In this scenario, the total flux at a given time $i$ in the filters 1 and 2 is calculated as  the sum of the intrinsic fluxes of the two stars, each magnified to different extents.
This is expressed as:
\begin{equation}
\begin{aligned}
&F_i\_1=F_A\_1*\mu_{i\_A}+F_B\_1*\mu_{i\_B}, \\
&F_i\_2=F_A\_2*\mu_{i\_A}+F_B\_2*\mu_{i\_B}.
\end{aligned}
\end{equation}
We then compute the magnitude difference $\Delta mag_\mathrm{AB}$ between the two filters (1 and 2) by finding the flux ratio $F_i\_1/F_i\_2$ at each time step.
Since each star undergoes a unique, time-varying magnification ($\mu_{i\_A}$ and $\mu_{i\_B}$), the magnification factors cannot cancel out in the flux ratio.
As a result, the flux ratio changes over time, leading to a temporal variation in the color-magnitude difference.

We propose that these multi-band magnitude differences could serve as a diagnostic tool for identifying binary star systems in observational data. 
To explore this phenomenon, we simulated multi-band light curves for the binary star systems described in Section 3 using the three broad-band filters (CBU, CBV, and CBI) of the CSST-MCI. 
Their wavelength coverage and transmission curve are illustrated in Fig.~\ref{fig:star_spectra_filters}.

Fig.~\ref{fig:Binary_lightcurves_3filters_M90M80} and Fig.~\ref{fig:Binary_lightcurves_3filters_M70M60} present the simulated multi-band light curves and color-magnitude differences for binary systems 1 and 2, respectively. Here, we present the case where the crossing angle is set to $\theta = \pi/2$.
As the stars in each binary system sequentially cross the micro-caustic and experience different magnifications, the magnitude difference between filters varies over time.
Specifically, for System 1 (Fig.\ref{fig:Binary_lightcurves_3filters_M90M80}), the magnitude difference in CBI-CBU fluctuations by approximately 0.5 magnitudes, where as for System 2 (Fig.\ref{fig:Binary_lightcurves_3filters_M70M60}), the fluctuations in CBI-CBU are about 0.15 magnitudes.  
These findings highlight the diagnostic potential of multi-band observations in identifying binary star systems and placing constraints on their properties.

\section{Summary}\label{sec:Summary}
In this study, we investigate the magnification patterns of high-redshift stars crossing micro-caustics, focusing on the distinctive light curve signatures produced by detached binary star systems.
Here, galaxy clusters act as natural gravitational lenses, magnifying these distant sources, while microlenses within the clusters further disrupt the smooth macro-caustic into a band of micro-caustics. 
Our findings provide a foundation for interpreting future observations of caustic-crossing events and distinguishing binary systems from single stars.

Using high-resolution magnification maps generated by the GPU-CAUSTIC, we validated the theoretical magnification curve for single stars crossing micro-caustics, confirming its applicability as a baseline for more complex systems. This theoretical foundation enabled us to model detached binary star systems and explore their unique light curve features under varying orbital parameters and caustic geometries. 
The massive star models we use in this work come from \texttt{PARSEC}. It has a detailed consideration on the mass loss of massive stars, which significantly influences the intrinsic magnitude and current mass  of the stars.

The results reveal that detached binary systems produce diverse magnification patterns, including overlapping peaks, plateau-like features, and temporal variations in color-magnitude differences. These characteristics depend on the binary stars’  orbital radii, crossing angles, crossing sequence, and intrinsic flux ratios, offering diagnostic tools for identifying binary systems in caustic-crossing events. 
Simulations of multi-band photometry using the CSST-MCI demonstrate that temporal fluctuations in color-magnitude differences, alongside light curve morphology, can effectively characterize detached binary stars in lensing events.

Future work will address more complex binary configurations such as  mass transfer effects in binary stars. These advancements will refine our understanding of the magnification signatures of binary systems and enhance their applicability to real-world observations, paving the way for deeper insights into high-redshift astrophysics.

\section*{Acknowledgements}
This work is supported by the National Natural Science Foundation of China (NSFC, No. 12403006, No.12333001,No.12288102, No.12203100, No. 12125303), the Jiangsu Province Distinguished Postdoctoral Fellowship Program, and the GHfund A (202407017555).
We acknowledge the science research grants from the China Manned
Space Project with NO.CMS-CSST-2021-A12 and No. CMS-CSST-2021-A08.

\section*{Data Availability}
This theoretical study does not generate any new data.



\bibliography{References}{}
\bibliographystyle{aasjournal}


\end{sloppypar}
\end{document}